\documentclass[preprint2]{aastex}
\input epsf.sty

\shorttitle{Emission components of PG1211+143}
\shortauthors{Janiuk et al.}

\begin{document}

\title{The nature of the emission components in the quasar/NLS1 PG1211+143}

\author{A. Janiuk and B. Czerny}
\affil{ Nicolaus Copernicus Astronomical Center, Bartycka 18,
            00-716 Warsaw, Poland}
\email{agnes@camk.edu.pl; bcz@camk.edu.pl} 

\author{G. M. Madejski}
\affil{Stanford Linear Accelerator Center,  2575 Sand Hill Road, Menlo
Park, CA 94025, USA}
\email{madejski@slac.stanford.edu}

\clearpage

\begin{abstract}

We present the study of the emission properties of the quasar 
PG1211+143, which belongs to the class of Narrow Line Seyfert 1 galaxies.
On the basis of observational data analyzed by us and collected
from the literature, we study the temporal 
and spectral variability of the source in
the optical/UV/X--ray bands and we propose a model that explains the
spectrum emitted in this broad energy range. In this model, 
the intrinsic emission
originating in the warm skin of the accretion disk is 
responsible for the spectral component that is dominant in the softest
X--ray range.  The shape of reflected spectrum as
well as Fe K line detected in hard X--rays require the reflecting
medium to be mildly ionized ($\xi \sim 500$). We identify this reflector
with the warm skin of the disk and we show that the heating of the skin
is consistent with the classical $\alpha P_{tot}$ prescription, while
$\alpha P_{gas}$ option is at least two orders of magnitude too low to
provide the required heating. We find that the mass of
the central black hole is relatively small ($M_{BH} \sim 10^{7}- 10^8 
M_{\odot}$), which is consistent with the Broad Line Region mapping
results and characteristic for NLS1 class. 

\end{abstract}


\clearpage
\section{Introduction}

The quasar PG1211+143 ($z=0.0809$;  Marziani et al. 1996) is one of 
the prototypes of AGN with particularly strong hard optical spectrum and 
profound steep soft X--ray emission (Elvis, Wilkes, \& Tananbaum 1985).
It has been frequently argued that the optical/UV and soft X--ray
emission form a single Big Blue Bump component which extends across the 
unobserved XUV band (Bechtold et al. 1987), and strongly dominates the 
bolometric luminosity. 

However, the nature of the soft X--ray emission -- in particular,
the spectral component that exhibits a very soft spectrum and often is
seen in many quasars and Seyfert 1 galaxies only below $\sim 1$ keV -- 
is still not determined unambiguously.  There are arguments
both in favor and against the origin of this emission component 
being the tail of the optical/UV Big Blue Bump.  This is because any
interpretation relies on extrapolating the spectrum over a
decade in frequency through the unobserved XUV band.

The strong arguments in favor of the Big Blue Bump (BBB) extending up to the
soft X--ray range are statistical in their nature. There is a strong, 
almost linear correlation between the luminosity of an object 
at 2500 \AA~ and the soft
X--ray luminosity in the ROSAT spectral band both in a large sample of
Seyfert 1 galaxies (Walter \& Fink 1993) and in radio quiet quasars
(Yuan et al. 1998) suggesting a relatively uniform spectral shape of 
the BBB. Spectral analysis of both high redshift and low redshift
quasars (Laor et al. 1997) allowed for a ``shrinking'' of the unobserved band 
and such spectra appear to cover smoothly the entire 
optical/UV/soft X--ray band.

Arguments against the soft X--ray emission being a direct tail of the
optical / UV component are also mostly statistical. It is rather surprising
from the theoretical point of view why the extension of the BBB should be the
same in all objects having a range of black hole masses and luminosities.
Instead, it was suggestive of some atomic processes (e.g. Czerny
\& \. Zycki 1994).  Also the normalized variability amplitude in the UV band 
is usually smaller than in the X--ray band (see e.g. Ulrich et
al. 1997).  A close look at the UV spectra of some sources frequently 
suggests a slight decline on $\nu F_{\nu}$ diagram in the far UV (Malkan 
1983;  Magdziarz et al. 1998 for NGC~5548).  
This does not really allow for this component to extend 
far enough into XUV range to make it observable in the 
soft X--ray band, even if
a certain amount of unsaturated Comptonization is taken into account.  

\begin{deluxetable}{lccr}
\tablenum{1}
\tablewidth{80 mm}
\tablecaption{X--ray observations
\label{tab1}}

%\begin{table}
%\caption{X--ray observations: Fluxes without Galactic absorption in $[10^{-12} $erg/s/cm$^2]$ .
%\label{tab1}}
%\begin{tabular}{llrr}

\tablehead{
\colhead{Instr.} &
\colhead{Date} &
\colhead{F(0.2 - 2)$^a$} &
\colhead{F(2 - 10)$^a$}
}

%instr. & date   &    F(0.2-2) &   F(2-10) \\

%%%%& & $[10^{-12} erg/s/cm^2]$ & $[10^{-12} erg/s/cm^2]$ \\
\startdata
EINSTEIN & 1979Dec05  &     52.1  &    7.3 \\
EINSTEIN & 1980Dec11  &     75.0  &   10.5   \\
EXOSAT   & 1985Jun13  &     59.7  &    9.9    \\
EXOSAT   & 1986Jan28  &     41.2  &    6.4 \\      
EXOSAT   & 1986Jan09  &     18.9  &    5.8  \\      
GINGA    & 1985May06  &            &    8.5    \\     
ROSATa   & 1991Dec17 &     16.1  &     \\
ROSATb   & 1992Jun17 &     21.9  &    \\
ROSATc   & 1992Jun28 &     30.6  &       \\
ROSATd   & 1992Jun28 &     34.2  &        \\   
ROSATe   & 1993Jun02 &     20.5  &     \\
ROSATf   & 1993Jun25 &     19.0  &    \\
ASCA     & 1993Jun03  &     8.5   &   3.7\\
RXTE      & 1997Aug16  &            &   12.9    \\
\enddata

\tablenotetext{a}{
Fluxes without Galactic absorption 
in units of $10^{-12}$ erg s$^{-1}$ cm$^{-2}$}
 
\end{deluxetable}

%\end{tabular}
%\end{table}

Detailed analysis of the broad band spectra of Seyfert 1 galaxy NGC~5548 
(Magdziarz et al. 1998)
led to a conclusion that the optical/UV spectrum is well modeled 
by an accretion disk, and the hard X--rays are reproduced by standard thermal 
Comptonization.  However, an additional component is needed to model the 
soft X--ray source: the most viable candidate here is another, optically 
thick Comptonizing medium possibly associated with the transition between the
disk and the hot plasma. This component would 
contain a major part of the source 
bolometric luminosity.  Such a model also supports the scenario including
a strong and extended Big Blue Bump with Comptonization playing a 
major role in formation of this component.  

In this paper we study in detail the quasar PG1211+143.  Our goal is 
to investigate (i) if a similar model also applies to this source, 
which is representative both for quasars and for Narrow Line Seyfert 1 
galaxies, (ii) if this model is unique and (iii) what is the physical 
basis and possible observational consequences of such a 
phenomenological model.  

The choice of object was motivated by the availability of considerable 
amount of observational data available, which span the 
last two decades.  The object can be considered as a typical, although 
rather faint quasar, with $\lambda F_{\lambda}$ at 5100 \AA~ about 
$5 \times 10^{44}$ erg s$^{-1}$ (Kaspi et al. 2000) for the Hubble 
constant $H_o = 75$ km s$^{-1}$ Mpc$^{-1}$.  The broad band  $\alpha _{ox}$ 
index measured between 2500 \AA~ and 2 keV is 1.16 (Corbin \&
Boroson 1996) while for a typical quasar, it is 1.5 (Yuan 
et al. 1998). The 
soft X--ray slope measured in the ROSAT band is $\sim 3.1$  
(Walter \& Fink 1993) while typical quasar value is 2.58 (Yuan et al. 
1998).   Its optical emission lines are narrow, with the FWHM of 
$H_{\beta}$ line is $1832\pm 81$ km s$^{-1}$ (Kaspi et al. 2000; see 
also Corbin \& Boroson 1996; Wilkes et al. 1999) implying a
classification of this object as a Narrow Line Seyfert 1 galaxy.  

In Section 2 we describe the observational data that we analyzed,
and how we supplement these with the data available from the 
literature. In Section 3 we give the results of the variability and 
spectral studies of PG1211+143. We discuss the physical processes 
that may lead to the observed behavior of the source in Section 4 
and summarize the results in Section 5.

\section{Observations and data analysis}

First spectral X--ray data for PG1211+143 were obtained from the 
EINSTEIN observatory (see Table~\ref{tab1}).  Strong soft X--ray 
emission seen in EINSTEIN was described by Bechtold et al. (1987), 
and it was subsequently reanalyzed by Elvis et al. (1991).  
We determine the 0.2 - 2 keV and 2 - 10 keV 
intrinsic fluxes (i.e. corrected for Galactic reddening) in the second 
observation from the fits provided by Bechtold et al. (1987). The 
quality of the first observation was much lower, so we determined 
the soft and hard fluxes in this data by rescaling the Bechtold et 
al. (1987) results with the IPC count rate ratios 0.695 between 
the two observations, as given by Elvis et al. (1991).

Subsequent observations were performed by EXOSAT on three occasions
(Elvis et al. 1991;  Saxton et al. 1993).  We determine the  2 - 10 
keV intrinsic fluxes through the spectral fitting of publicly
available ME data to a simple broken power law model with fixed 
galactic absorption. The 0.2 - 2 keV flux was calculated from the 
results of spectral fitting to LE data presented in Saxton et al. (1993). 

On one occasion the source was observed by GINGA. We take the 2 - 10 keV flux
from Lawson \& Turner (1997).

PG1211+143 has been observed six times by the ROSAT satellite with the
PSPC detector.  One of the ROSAT pointings was simultaneous with 
ASCA observations.  The ROSAT and ASCA spectral data were reduced 
using the standard software package, where for ASCA, we used data 
from all four detectors (SIS0/1 and GIS2/3) fitted simultaneously.  
Preliminary results based on these data were 
presented by Yaqoob et al. (1994) but we repeat their analysis, 
using the updated, current version of the ASCA response matrices. The ASCA 
observation was sufficiently long (88 ksec) to attempt a study of the 
light curve within this single data set, so we could also determine 
the variability on short time scales.

The Rossi X--ray Timing Explorer (RXTE) 
observations of PG1211+143 are new, conducted as a part of 
optical / X--ray monitoring program.  The object was monitored roughly 
every 5 days for 6 months;  the results of this program 
will be reported elsewhere (Netzer, Madejski, Kaspi, et al., in 
preparation).  Here, we use the spectrum derived from the 
summed RXTE Proportional Counter Array (PCA) data
analyzed in a standard manner (including the background subtraction), 
using the {\sc ftool} script {\sc rex}.   The source was detected to at least 
12 keV, beyond which the signal - to - noise ratio was becoming very low.  
There was no significant detection with the 
RXTE's High Energy X--ray Timing Experiment (HEXTE).  

All analysis of the spectral X--ray data was done using the XSPEC 
software package, version 10.0 (Arnaud 1996).  We adopt a redshift 
0.0809 to this source after Marziani et al. (1996) instead of 0.085 
used in most previous papers (e.g. Bechtold et al. 1987;  Yaqoob et
al. 1994) since the new determination seems to be convincing.
Whenever we fix the Galactic column density, we adopt the value 
of $N_{H}=2.83 \times 10^{20} $cm$^{-2}$ after Elvis, Wilkes, \& Lockman 
(1989) although another value ($N_{H}=2.1 \times 10^{20} $cm$^{-2}$) was 
used by Lawson \& Turner (1997);  we note that the exact value of
$N_{H}$ makes essentially no difference in our analysis.  

The optical/UV spectrum of this source was determined by 
Bechtold et al. (1987). We use the broad band data of 
Elvis et al. (1994) and we compare them with the HST observations 
of Bechtold et al. (2000), performed on April 13 and 16, 1991.

\section{Results} 
\label{sect:results}

\subsection{Time variability}

\subsubsection{Short timescales in the X--ray band}
\label{sect:short}

\begin{figure}
\epsfxsize = 80 mm 
\epsfbox{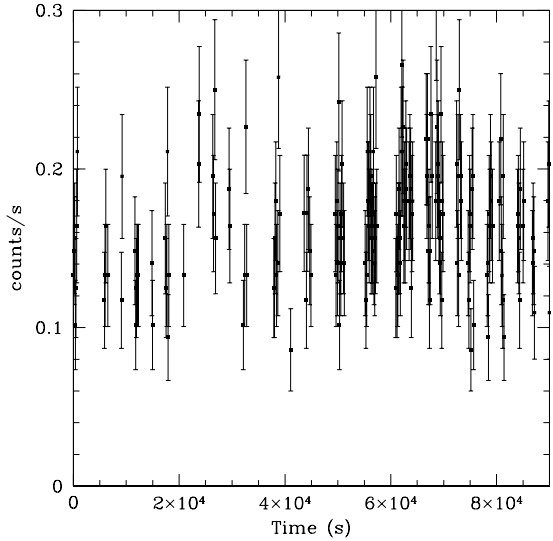}
\caption{The ASCA GIS-2 lightcurve in 2 - 10 keV energy band.
\label{fig:ASCA_lc}}
\end{figure}

We study the short timescale variability in the hard and soft X--ray 
bands on the basis of the longest sequence from ASCA (June 3 - 4, 1993, $8.7 
\times 10^4$ s), and accompanying ROSAT (June 2 - 4, 1993, $4 \times 10^4$ s).
We analyzed the ASCA lightcurve binned into 128 s bins for two SIS and 
two GIS instruments separately, in the entire energy band. As an example,
we show the GIS-2 lightcurve in Figure~\ref{fig:ASCA_lc}.

The source is clearly variable in both energy bands.  The rms amplitude in
ASCA and ROSAT bands are given in Table~\ref{tab2}.  Such values are 
comparable to those measured in X--rays in other AGN, e.g. 16\% in 
MCG--6-30-15 on timescales shorter than $10^6$ s (Nowak \& Chiang 2000). 
The count rate in ASCA is dominated by 1 - 2 keV energy band, where both 
the contribution of the soft X--ray excess and hard X--ray power law 
are important. We therefore compared the rms value derived above 
with the rms value measured from the light curve for photons 
above 2 keV. We used only the data from GIS instruments as 
those data are more reliable for time-series 
analysis than SIS data for PG 1211+143. This is because the PG 1211+143 
ASCA observation was conducted in a 4-CCD mode of the SIS instrument,
and the source image fell close to the gap between the CCD chips.
Since the satellite attitude can change somewhat during the observation,
some unknown fraction of photons may fall in the space between the chips,
resulting in an apparent but erroneous variability. Both GIS2 light curves
(whole energy band, and above 2 keV) were rebinned into 512 s bins
to increase statistics;  the two observations yield rms values of 
0.095 and 0.171, respectively.  This suggests that the observed 
variations at the time scales of minutes-hours are predominantly due 
to the hard X--ray power law.  This is confirmed by the fact that rms 
value measured in the ROSAT band is much smaller during 
the 40 ksec observation than in the ASCA data.  

The data from ASCA and ROSAT satellites are exactly simultaneous;  
nonetheless, the presence of gaps due to the Earth occultation
of the source makes an attempt to use the Fourier analysis rather 
difficult, although there are methods which allow to overcome this 
problem for high quality observations (e.g. Done et al. 1992a). 
However, even the available ASCA data are not of sufficient quality to 
perform this kind of analysis.  Therefore, here we adopt a 
complementary method based on a nonlinear prediction, which is 
described in Czerny \& Lehto (1997). We analyzed the light curve
binned into 128 s bins for two SIS and two GIS instruments separately, 
in the entire energy band.  Such analysis reveals that the shape of 
the power density spectrum (hereafter PDS) is basically described 
as a power law, with the slope 1.5. Errors in determination 
of those values are substantial, and systematic errors dominate, 
since the existence of the signal in the data is marginal (the
correlation coefficient itself is 0.62 and 0.67, i.e. below 0.8, which 
would be necessary for reliable results). 

Such a value of the PDS slope is typical for Seyfert galaxies at 
high frequencies. Yaqoob (1997) obtained the high frequency slope 
of MCG--6-30-15 of 1.4, Edelson \& Nandra (1999) determined the slope
to be 1.74 in NGC~3516 at frequencies above $\sim 10^{-6}$ Hz, and 
Nowak \& Chiang (2000) as well as Chiang et al. (2000) give the slope
of 1.0 at intermediate frequencies and a slope of 2.0 at high
frequencies in MCG--6-30-15 and NGC~5548, respectively.

\begin{figure}
\epsfxsize = 80 mm 
\epsfbox{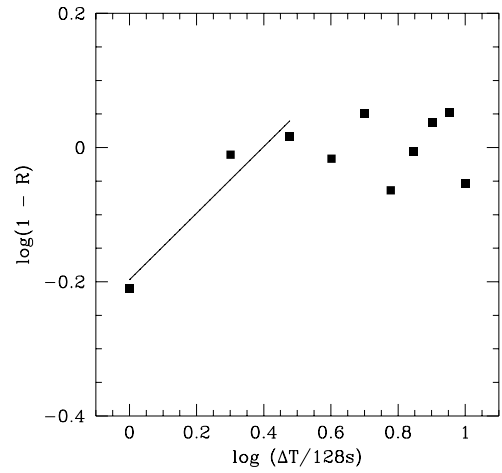}
\caption{The dependence of the decay of correlations with the time 
step in the ASCA lightcurve. First three points were used to determine 
the slope (solid line) which directly translates into the slope of 
the power density spectrum.
\label{fig1}}
\end{figure}

We conclude that the variability properties of PG1211+143 are rather typical
of a Seyfert galaxy. Unfortunately, neither quasars nor Narrow Line Seyfert 
1 galaxies were studied with equal accuracy as normal Seyfert 1
galaxies, so it is not known whether 
there are any systematic differences in variability
apart from scaling with mass (e.g. Leighly 1999;  Piasecki et
al. 2000). 
Results for PG1211+143 may indicate that the variability pattern is universal.

\begin{deluxetable}{lcccr}
\tablenum{2}
\tablewidth{80 mm}
\tablecaption{Variability properties of PG1211+143
\label{tab2}}

%begin{table*}
%\caption{Variability properties of PG1211+143.
%\label{tab2}}
%\begin{tabular}{ccccc}

\tablehead{
\colhead{Timescale} &
\colhead{F(0.2 - 2)$^a$} &
\colhead{rms}&
\colhead{F(2 - 10)$^a$}&
\colhead{rms}
}

%%timescale   &  F(0.2 - 2) & rms &  F(2 - 10) & rms \\

%%%s & $[10^{-12} erg/s/cm^2]$& & $[10^{-12} erg/s/cm^2]$ &\\
\startdata
$3.3 \times 10^4$ & 17.9  &  0  \\
$8.7 \times 10^4$ &       &       &    3.96   &  0.09 \\
$1.7 \times 10^5$ & 18.2  &  0.06 &       &  \\
$4.8 \times 10^7$ & 23.7  &  0.25 \\
$4.3 \times 10^8$ & 36.5  &  0.62   &  7.44   & 0.30  \\
$5.6 \times 10^8$ &       &         &  8.12   & 0.34  \\
\enddata
\tablenotetext{a}{
Fluxes in units of $10^{-12}$ erg s$^{-1}$ cm$^{-2}$}
\end{deluxetable}
%\end{tabular}
%\end{table*}

\subsubsection{Trends of long-term variability in the X--ray band}
\label{sect:long}

Available archival data for PG1211+143 include six ROSAT pointings 
as well as historical observations from EINSTEIN, EXOSAT, GINGA, and
ASCA;  with those, we can also study the long time variability
patterns of the source.  The ROSAT observations cover 
a period of 1.5 years, allowing for a reliable measurement with 
the same instrument.  We determine the observed fluxes
in 0.2 - 2 keV band assuming a simple spectral model of a broken power law,
with the second (hard X--ray) power law photon index fixed at 2.0 and the
Galactic absorption fixed at the assumed value as given above.  
The ROSAT sequences $c$ and
$d$ obtained on the same day were both fitted separately and combined
(see Table~\ref{tab3}). Since the errors in separate fits are large, 
we use only the combined set in further analysis. All flux
measurements are summarized in Table~\ref{tab1}.

The variability during the period of ROSAT observations is moderate, roughly 
by a factor 2. More precisely, the rms value measured on 
the timescale of $4.8 \times 10^7$ s is only 0.25, which is 
not much higher than the rms variability on 
short time scales, within a single ASCA data set. 

However, at the timescales of years the trend reverses, and the soft
X--ray variability amplitude becomes higher than the hard X--ray 
variability on the same time scale. Analyzing the entire period of 
twenty years of observations, we measure the rms in 0.2 - 2 keV band
which is higher than in 2 - 10 keV band (see Table~\ref{tab2}).  
It is consistent with the trends shown by the Big Blue Bump in other AGN 
(e.g. Clavel et al. 1992 for NGC~5548 and Clavel, Wamsteker \& Glass 1989 for 
Fairall~9).

\begin{figure}
\epsfxsize = 80 mm 
\epsfbox{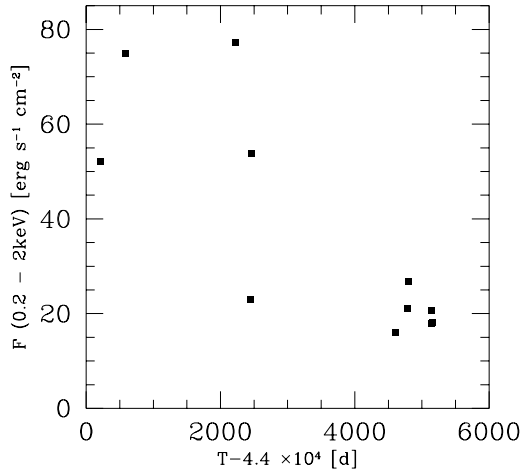}
\caption{The time dependence of the soft 0.2 - 2 keV flux from instruments: EINSTEIN, EXOSAT, ASCA/ROSAT and ROSAT.
\label{fig2}}
\end{figure}

The most dramatic change happened during the end of 1985/beginning of 
1986 but this period is covered by only 3 EXOSAT  observations (see 
Figure~\ref{fig2}). Large changes in the soft X--ray emission are 
accompanied by the changes in the hard X--ray flux but not with 
comparable amplitude (see Figure~\ref{fig1}).

Since the analysis of ASCA data allows us to estimate the slope of the 
PDS at high frequencies, we can reconstruct the total PDS from the 
values of rms on time scales of $8 \times 10^4$ s and $5 \times 10^8$
s. We assume the slope of 1.5 at high frequencies and a slope 0.0
below  certain frequency 
$\nu_o$. Using the integral conditions connecting the PDS and rms at 
a given frequency (see e.g. Czerny, Schwarzenberg-Czerny, \& Loska 
1999) we obtain the flattening frequency $\nu_o = 8.4 \times 10^{-7}$
Hz, corresponding to a timescale of $1.2 \times 10^6$ s. This is 
again comparable to the results for most Seyfert galaxies: the flat 
part of the PDS starts at timescales $\sim 10^6$ s for MCG--6-30-15 
(Nowak \& Chiang 2000), $\sim 10^7$ s for NGC~5548 (Chiang et
al. 2000) and $\sim 10^8$ s for NGC~3516 (Edelson \& Nandra 1999).
 
\begin{figure}
\epsfxsize = 80 mm 
\epsfbox{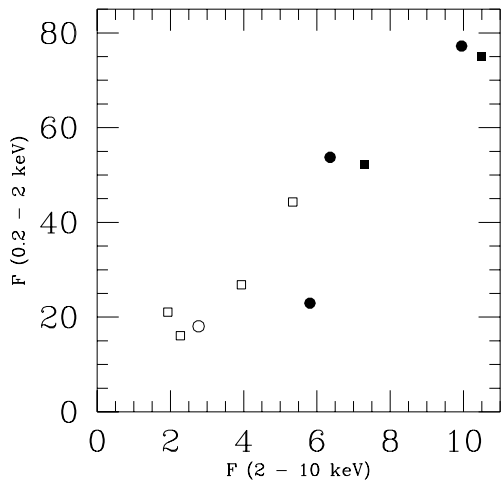}
\caption{The soft 0.2 - 2 keV flux as a function of 2 - 10 KeV hard 
X--ray flux from various instruments: EINSTEIN (filled squares), 
EXOSAT (filled circles), ASCA/ROSAT (open circle), and ROSAT (open squares).
\label{fig1}}
\end{figure}

\subsubsection{Black hole mass determination from variability} 
\label{sect:masa}

X--ray variability time scales have been successfully used to estimate 
the mass of the central black hole in active galaxies (e.g. Barr \& 
Mushotzky 1986;  Wandel \& Mushotzky 1986), although a more robust
method is the measurement of features in the PDS (see, e.g., Edelson
\& Nandra 1999).  Convenient and powerful method based on the
normalization of the high frequency tail of the power density spectrum was
recently developed by Hayashida et al. (1998) and Hayashida (2000). 
Here we use a modified version of this method (Piasecki et al. 2000) 
which is based on comparison of the shape of the PDS 
with independent mass determination for the Seyfert 1 
NGC~5548, an AGN;  for the purpose of the mass estimate, 
this is likely to be more appropriate than scaling the 
normalization of Cyg X-1, an accreting binary system, 
used in Hayashida et al. 1998.  
The data discussed in previous section are not of sufficient quality 
to perform Fourier analysis. However, having the rms determined for a 
few time scales, we can reconstruct the power density spectrum
assuming its shape in advance and using the integral connection between 
the power density spectrum and rms (see e.g. Czerny et al. 1999).

Here we assume that the PDS is flat below certain frequency 
and has a slope 1.5 above it, as determined in
Section~\ref{sect:short}.  As data points for this method, we 
use the GIS rms measurement for the $8.7 \times 10^4$ s time scale 
and multi-instrument measurements on the $ 4.3 \times 10^8$ s time 
scales. Extension of the measurements afforded by the RXTE
observations only confirms that the break frequency is at timescales
below $\sim 10^8$ s;  if this timescale is above $\sim 10^8$ s, 
the rms would saturate.  With this, we determine the break frequency 
to be at $ 2 \times 10^{-7}$ Hz, corresponding to timescales of 
$5 \times 10^6$ s. The error on this value is large, since 
the uncertainty of the GIS rms measurement is already a factor 5.

This simple approach gives the value of the black hole mass 
log$M$ = 7.0 $\pm$ 0.7 [$M_{\odot}$], if the error in determination 
of the rms in GIS data (0.3) is taken into account. The mass of 
PG1211+143 is thus definitely rather small for a quasar, and more 
typical for Seyfert galaxies. High luminosity of the object has to be 
related to a relatively high accretion rate, close to the Eddington value.  
Small mass is consistent with the measurements of Kaspi et al.
(2000) based on the properties of the Broad Line Region: 
$M = 4.05 ^{+0.96}_{-1.21} \times 10^7 M_{\odot}$, $L/L_{Edd} = 0.29$. 

%\begin{table*}
%\caption{Fits to ROSAT data: model bknpower; 0.2 - 2 keV fluxes without 
%Galactic absorption.
%\label{tab3}}
%\begin{tabular}{cccccccc}

\begin{deluxetable}{lcccccr}
\tablenum{3}
\tablecaption{Fits of broken powerlaw model to ROSAT data
\label{tab3}}
\tablewidth{120 mm}
\tablehead{
\colhead{Data} &
\colhead{$N_{H}$} &
\colhead{$\Gamma_{1}$} &
\colhead{$E_{break}$} &
\colhead{$\Gamma_{2}$} &
\colhead{ $\chi^2/d.o.f.$} &
\colhead{Flux(0.2 - 2)$^a$} 
}

%                                                           (no abs)
%data  &$N_{H}$    &$\Gamma_{1}$          &Ebreak                                      & $\Gamma_{2}$ &  $\chi^2/d.o.f.$ & Flux.2-2\\

\startdata
a  &3.45$^{+0.44}_{-0.35}$ &3.09$^{+0.17}_{-0.14}$ 
  &1.83$^{+\infty}_{-0.50} $ &2.0 &39.0/36 & 16.13 \\
b  &3.46$^{+0.46}_{-0.42}$ &3.33$^{+0.18}_{-0.17}$ 
  &$>$ 2.0$^b$ &2.0 &39.6/36 & 21.94 \\
c  &3.36$^{+0.65}_{-0.60}$ &3.22$^{+0.24}_{-0.23}$ 
  &$>$2.0 &2.0 &30.3/36 & 30.61 \\
d  &3.33$^{+0.79}_{-0.70}$ &3.17$^{+0.30}_{-0.28}$ 
  &1.41$^{+\infty}_{-0.19} $ &2.0 &48.2/36 & 34.25 \\
cd  &3.20$^{+0.44}_{-0.44}$ &3.11$^{+0.17}_{-0.16}$ 
  &1.67$^{+\infty}_{-0.31}$  &2.0 &48.5/36 & 30.92 \\
e  &3.65$^{+0.43}_{-0.40}$ &3.38$^{+0.16}_{-0.15}$ 
  &$>$2.0 &2.0 &53.2/36 & 20.53 \\
f  &3.50$^{+0.49}_{-0.33}$ &3.18$^{+0.17}_{-0.15}$ 
  &1.78$^{+\infty}_{-0.35} $ &2.0 &62.3/36 & 18.99 \\
\enddata

%\end{tabular}

\tablenotetext{a}{Fluxes without Galactic absorption}
\tablenotetext{b}{Energy break beyond the ROSAT band}
%\end{table*}
\end{deluxetable}

\begin{deluxetable}{lcccccr}
\tablenum{4}
\tablecaption{Fits to combined ROSAT data with variable normalization
\label{tab4}}
\tablewidth{140 mm}
\tablehead{
\colhead{Model} &
\colhead{Par1$^a$} &
\colhead{Par2$^b$} &
\colhead{Par3$^c$} &
\colhead{Par4$^d$} &
\colhead{Par5$^e$} &
\colhead{$\chi^2/d.o.f.$}
}
%\begin{table*}
%\caption{Fits to combined ROSAT data with variable normalization;
%fluxes without 
%Galactic absorption. Par1 is the galactic 
%absorption in units $10^{20} $cm$^{-2}$, par2 is the soft spectral
%index (models A and B) or the soft photon 
%temperature in keV (models C and D), 
%par3 is the break energy (model A) or the electron 
%temperature (model D) and par4 is the hard X--ray spectral 
%index (model A) or the optical depth (model D).
%\label{tab4}}
%\begin{tabular}{cccccccc}
%
%                                                           (no abs)
%model  & par1    &   par2    &par3  & par4  & $\chi^2/d.o.f.$ \\

\startdata
A bknpower  &3.54$^{+0.18}_{-0.17}$ &3.25$^{+0.07}_{-0.06}$    
& $> 1.75$  & 2.0 &     &306/192 \\
B power     &3.53$^{+0.18}_{-0.17}$ &3.25$^{+0.07}_{-0.06}$    
&         &     &      & 306/191 \\
C bbody+power    &1.31$^{+0.13}_{-0.12}$ &0.107$^{+0.003}_{-0.003}$ 
&       &2.0 &       & 331/191 \\
D comptt       &2.83                   & $< 0.020$  &
0.39$^{+0.07}_{-0.06}$&        & 10.4$^{+0.8}_{-1.0}$ &294/191 \\
E comptt+power &2.83                   & $< 0.017$  &
0.24$^{+0.11}_{-0.08}$& 2.0    &14.6$^{+4.5}_{-3.1}$ &290/190 \\

\enddata

\tablenotetext{a}{Galactic absorption in units $10^{20} $cm$^{-2}$}
\tablenotetext{b}{Soft spectral index (models A and B) or the soft
photon temperature in keV (models C and D)}
\tablenotetext{c}{Break energy (model A) or the electron temperature
(model D)}
\tablenotetext{d}{Hard X spectral index (models A, C and E)}
\tablenotetext{e}{Optical depth (models D and E)}
%\end{tabular}
%\end{table*}
\end{deluxetable}

\subsubsection{Spectral variability in the soft X--ray band}

\begin{figure}
\epsfxsize = 80 mm 
\epsfbox{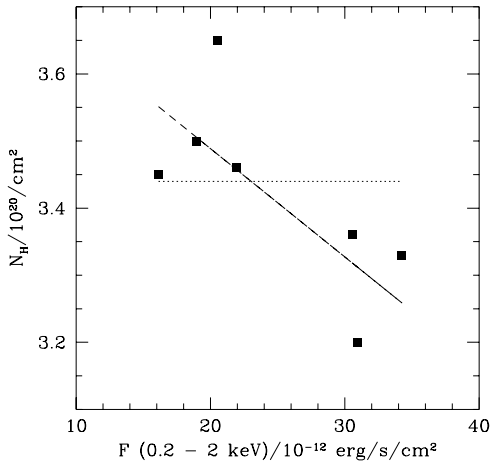}
\caption{The relation between the fitted value of $N_H$ 
and the soft 0.2 - 2 keV flux in the ROSAT data. Dotted 
line represents the best fit constant value. Dashed 
line represents the best linear fit. Errors in  $N_H$ 
are very large ($\sim 0.4$; see Table~\ref{tab3}).
\label{figNh}}
\end{figure}

\begin{figure}
\epsfxsize = 80 mm 
\epsfbox{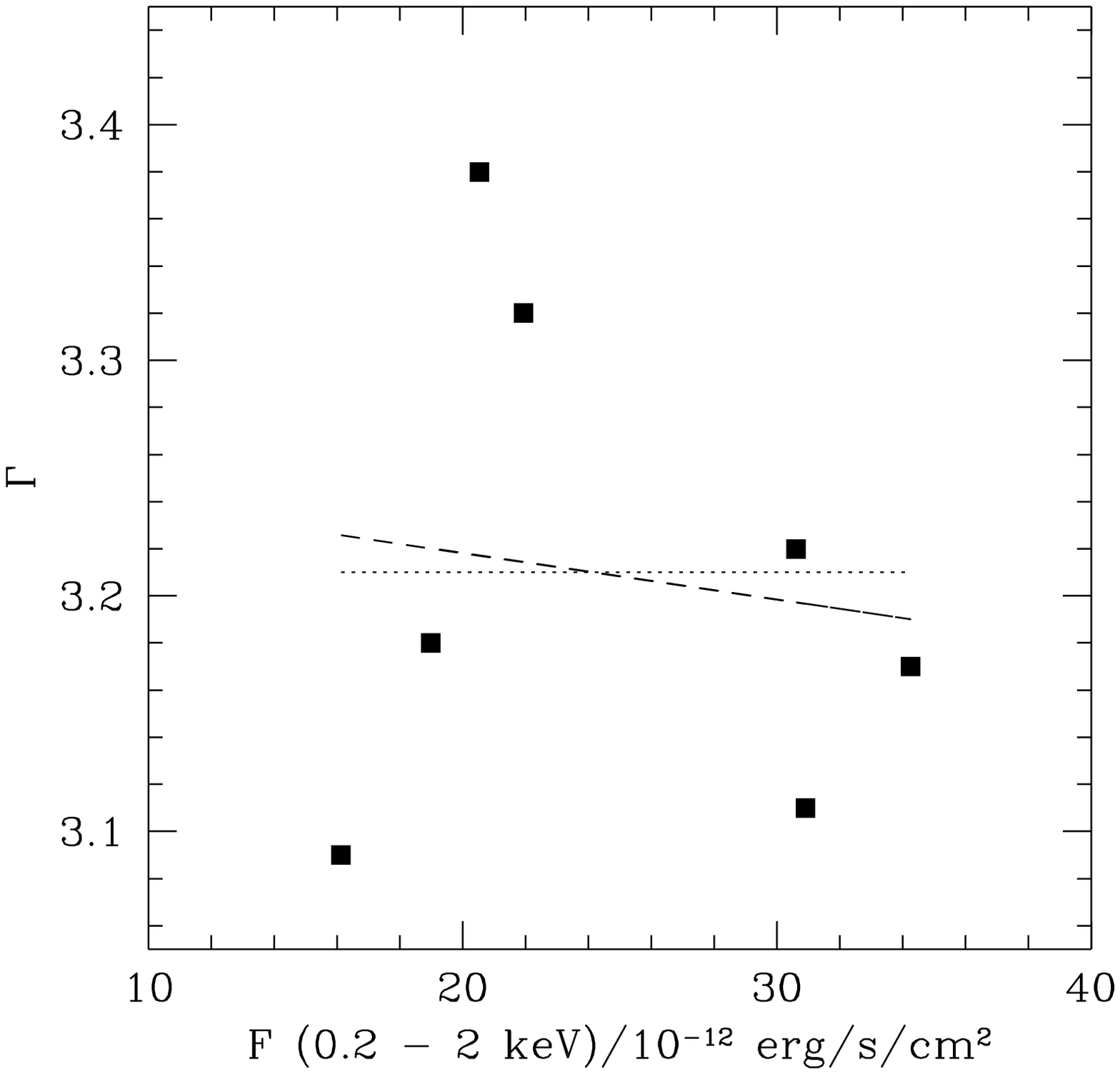}
\caption{The relation between the value of photon index $\Gamma$ 
and the soft 0.2 - 2
keV flux in the ROSAT data. Dotted line represents the best fit 
constant value. Dashed line represents the best linear fit.  
Errors in $\Gamma$ are large ($\sim 0.2$; see Table~\ref{tab3}).
\label{figGamma}}
\end{figure}

While the overall X--ray spectral variability as discussed above can 
be interpreted as a relative change of the intensity of the soft
vs. the hard component, we also aim to determine if the soft component 
{\sl alone} of the total spectrum of PG1211+143 has a variable shape.  
To this end, we fit the ROSAT data with a spectral model consisting 
of a cold absorber and a broken power law. The results are shown in 
Table~\ref{tab3}. We also searched for the presence of an ionized 
absorber in this source by adding an {\sc absori} XSPEC 
model in the case of the data set $f$, which shows the worst $\chi^2$. 
The fit was only marginally improved ($\Delta \chi^2 = 7.0$, 
with 34 d.o.f. instead of 36) and the best fit value of the column 
of the ionized gas was $ 5.5 \times 10^{21}$ cm$^{-2}$ (between 1.5 
and 13.0), with a temperature $kT \sim 1 \times 10^6$ K and 
$\xi \sim 20$, where both $T$ and $\xi$ are only weakly constrained. 

The data sets clearly show the variability of the flux of the soft
component, as discussed in the previous section.  Those variations, 
however, are not accompanied by significant changes in the spectral 
shape.  There is a slight trend in the variations of the model value 
of $N_H$ with the total flux. However, no such trend is visible in 
the photon index.  Formal analysis of the $N_H - F_{\nu}$ relation 
shows that the two parameters are best represented by a linear fit 
shown in Fig.~\ref{figNh} but a constant line representation is also 
acceptable ($N_H=3.44 \pm 0.17$, $\chi^2 = 0.60$ for 6 d.o.f.) since 
the errors in determination of $N_H$ are very large (see Table~\ref{tab3}).
Similar analysis of the $\Gamma - F_{\nu}$ trend confirms the visual 
impression that the correlation between the two quantities is 
insignificant - and a constant line fit gives $ \Gamma = 3.21 \pm
0.07$ ($\chi^2 = 2.6$ for 6 d.o.f.). With this, we conclude that 
the shape of the soft X--ray component does not vary, within the 
accuracy of the available data.

\subsubsection{Optical/UV variability}

PG1211+143 was extensively monitored in the optical band in the period
1966 - 1982, but no variability was found within the accuracy of
observations (Barbieri \& Romano 1984).  Bechtold et al. (1987) found 
the UV flux from IUE measurements in 1981 and 1982 that was 
systematically lower than the trend suggested by the optical data, 
although with a similar slope. Wit this, they grey-shifted the UV data by +0.27 
in logarithmic scale to match the the optical data. More recent 
HST observations (Dobrzycki, private communication) give exactly the 
same UV flux as the grey-shifted value of Bechtold et al. (1987)
(-25.68 of Dobrzycki at 1387 \AA~ vs. -25.69 of Bechtold at 1370-1424 
\AA~ in the same logarithmic units), making it difficult to
distinguish the true variability from an instrumental effect.  The
question was finally settled by the  monitoring performed over a 7 1/2
year time span by Kaspi et al. (2000). New optical data clearly showed 
the variability by a factor of two, and the normalized
variability amplitude at 5100 \AA~ was determined to be 16.2\%. 

\subsection{Spectral fitting of the broad-band data}

Here, we perform spectral fits using simple phenomenological models as
well as more realistic, physical models to individual data sets, 
and then make an attempt to fit the broad-band data simultaneously.  
We note that despite the changes in the normalization of each component, the 
{\sl shape} of each spectral component varies very weakly with time.  
With this, we attempt to describe this shape more accurately 
by using the combined data for each component.  We assume all model 
parameters are equal in all data sets apart from normalization.  

\subsubsection{ROSAT data}

Firstly, we analyze the combined ROSAT PSPC data, which we 
fit simultaneously using the same spectral model with variable normalization.
We fit the following
phenomenological models:  broken power law with a variable break 
energy but with fixed hard photon index of 2.0 (model A); 
a single power law (model B);  and black body with an underlying 
power law of photon index 2.0 (model C).  We also fit more physical 
models, namely:  thermal Comptonization model ({\sc comptt}) of 
Titarchuk (1994) parameterized by the soft photon temperature, hot 
plasma temperature and its optical depth (model D).  The best fitting 
results are presented in Table \ref{tab4}.  

The last model clearly provides the best description of all the ROSAT 
data, although the $\chi^2$ is still quite high. The soft photon 
temperature has only an upper limit of 0.034 keV, or $4 \times 10^{5}$ 
K. Such temperature is consistent with the effective temperature 
of a standard accretion disk around a black hole with the mass inferred 
by us above.  The hot plasma temperature is 
relatively low, $kT_{e} \sim 0.4$ keV or $5 \times 10^6$ K, much 
lower than the plasma responsible for the formation of the hard 
X--ray emission, and the optical depth of the plasma is very high, 
$\tau \sim 10$. The additional power law tail added in the model does not
improve this fit substantially, since the ROSAT data cannot constrain the 
spectral slope in hard X--rays.

We also tried to improve the quality of fits to 
the spectrum by including the in the model the O{\sc VIII} emission
line. The model parameters (model E in table  \ref{tab4}) remained basically 
unchanged and the line energy was fitted to 653 eV. However, the fit was not 
improved ($\chi^2/d.o.f.$=287/189), and the line equivalent width
could not be constrained ($EW=32^{+20}_{-32}$ eV). 

\subsubsection{Combined ROSAT/ASCA data - model without reflection}

\begin{figure}
\epsfxsize = 80 mm 
\epsfbox[10 100 600 700]{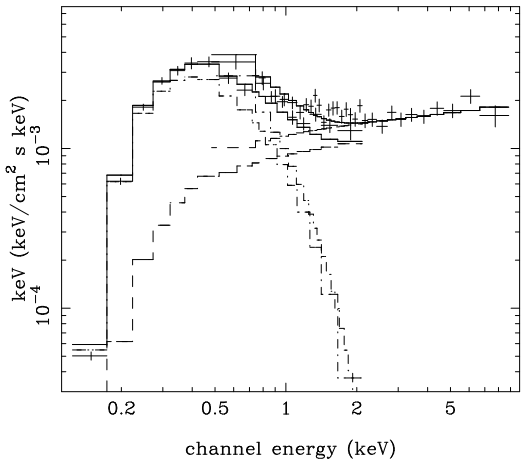}
\caption{The model of the combined ROSAT/ASCA data (see Table~\ref{fits}).
\label{fig:ROSAT/ASCA}}
\end{figure}

We apply our best model inferred above 
to ASCA/ROSAT data, which include all ROSAT 
pointings;  in the fits we use data from all four ASCA detectors, 
as described previously.  
We supplement the {\sc comptt} component with an underlying 
hard X--ray power law component characterized by its photon index, and
assume fixed Galactic absorption with $N_H = 2.8 \times 10^{20}
$cm$^{-2}$.  The results are given in Table ~\ref{fits}, and the spectra 
are displayed in Figure~\ref{fig:ROSAT/ASCA}. 

The fit to the data in the entire ROSAT/ASCA energy band is acceptable, and 
the Comptonizing plasma parameters are only slightly modified such
that the Comptonizning plasma temperature is now lower, and the
optical depth increased by a factor of 2. However, the basic picture 
remains qualitatively unchanged. The hard X--ray index 
is close to typical values for Seyfert galaxies and somewhat 
smaller than typically measured in quasars.
In this fit the reflection component was neglected.

The results for the combined ROSAT/ASCA data differ slightly from the
results obtained on the basis of the ASCA data alone by Vaughan et
al. (1999). The model used by Vaughan et al. is much simpler (power
law + black body).  Nevertheless, it is clear that the slope of the 
power law in our fit is flatter than in theirs (1.88 versus 2.07) and 
the soft X--ray excess is relatively stronger - the ratio of the 
luminosity of the soft X--ray excess to that of the hard power law 
in the 0.6 - 10 keV ASCA band is 0.28 in our fits and 0.17 in theirs.
This difference is probably caused by our inclusion of the ROSAT data and 
can be attributed to residual calibration uncertainties between ROSAT
PSPC and ASCA, resulting in different 
spectral slopes (see Iwasawa et al. 1999, describing the analysis of
nearly-simultaneous ROSAT PSPC and ASCA data for NGC~5548). 

\begin{deluxetable}{lcccr}
\tablenum{5}
\tablecaption{Results of simultaneous spectral fitting to ASCA and
ROSAT data
\label{fits}}
\tablewidth{80 mm}
\tablehead{
\colhead{$\Gamma$} &
\colhead{ $kT_{soft}^a$} &
\colhead{ $kT^a$} &
\colhead{ $\tau$} &
\colhead{ $\chi^{2}_{\nu}$} 
}

%\begin{table*}
%\caption{Results of simultaneous spectral fitting to ASCA and ROSAT
%data. 
%The model consists of the Comptonized black body, power law and Galactic absorption (fixed value).
%\label{fits}}
%%%\begin{tabular}{ccccc}

%$\Gamma$ & $kT_{soft}$  [keV]  & $kT$ [keV]  & $\tau$  & $\chi^{2}_{\nu}$ \\

\startdata
$1.88 \pm 0.05$  & $<0.02$ & $0.15 \pm 0.01$ & $21 \pm 2$ & 1.18\\
\enddata
\tablenotetext{a}{Temperature in keV}
\tablecomments{The model consists of the Comptonized black body, power law and Galactic absorption (fixed value).}
%\end{tabular}
%\end{table*}
\end{deluxetable}

\subsubsection{ASCA data - model with reflection}

The spectral fitting of 
the ASCA data alone were presented by Vaughan et al. (1999)
for the case of a simple model consisting of a power law, a black body and a 
Gaussian modeling the iron $K{\alpha}$ line. While the description of the
soft X--ray excess with a simple black body component is justified in
the ASCA data - these data 
are not sensitive to the shape of the soft excess - the presence of X--ray 
reflection in the data was not adequately studied by these authors, although
the detection of the iron $K_{\alpha}$ line in the ASCA data (Yaqoob et al.
1994;  Vaughan et al. 1999) clearly shows its importance.

Current models for the structure of AGN interpret the presence of 
the iron $K_{\alpha}$ line resulting from the irradiation of the
accretion disk surface by hard X--ray emission.  Such a fluorescent line is 
accompanied by the presence of a reflection component which is 
not an independent spectral feature (see, e.g., George \& Fabian
1991).  Therefore, we considered the properties of the entire reflection 
component, i.e. reflection hump plus the iron line (\. Zycki, 
Done, \& Smith 1997;  1998).  The reprocessed spectrum is parameterized 
by the ionization parameter $\xi=L_{X}/n_{e} r^{2}$ (Done et al. 1992b) 
and the reflection amplitude $R=\Omega/2 \pi$, where $\Omega$ is the 
solid angle subtended by the reprocessor as viewed from the X--ray
source.  We applied two versions of this model: one that does not contain 
the effect of relativistic smearing and another that includes this 
effect. The second model formally contains additional free parameters, but
some of these parameters of the model were always held fixed.  In our
case, the power law index of the radial distribution of the incident flux
was fixed at -3.0, the entire disk surface was assumed to be a reflector 
(i.e. $R_{in} = 6 R_g$, $R_{out}= 10^3 R_g$, where $R_G = GM/c^2$), and the inclination 
angle was assumed to be $30^o$.  The effect of these studies is shown in
Table~\ref{tabasca}. 

Formal minimum $\chi^{2}$ is for the relativistically smeared neutral 
reflection, but the normalization of the reflection component is surprisingly
high.  The assumption about the model adopted for the soft X--ray 
excess was not essential - fitting the {\sc comptt} model instead of
a black body led to the same values of relativistic reflection. 
If the amount of reflection is fixed at 1.0 the reflection is possibly 
ionized ($\xi = 400^{+1000}_{-400}$) but $\xi$ is not well
constrained.  If the reflection is assumed to be neutral, the amplitude of 
reflection is high ($R = 2.8^{+1.1}_{-0.9}$).  The quality of the 
fits is the same as before, and the $\chi^2$ for the second option is 
lower by 3.0.  

\begin{deluxetable}{lccccr}
\tablenum{6}
\tablecaption{Results of fitting the reflection component to ASCA data.
\label{tabasca}}
\tablewidth{120 mm}
\tablehead{
\colhead{Refl} &
\colhead{$\Gamma$} &
\colhead{ $kT_{bb}^a$} &
\colhead{ $\xi$} &
\colhead{ $R$} &
\colhead{ $\chi^{2}/d.o.f.$} 
}

% \begin{table*}
%\caption{Results of fitting the reflection component to ASCA data. The model consists of the black body, power law, Galactic absorption (fixed value) and
%the reflection component with a an iron line.
%\label{tabasca}}
%\begin{tabular}{cccccc}
%Refl& $\Gamma$ & $kT_{bb}$  [eV]  & $\xi$  & R  & $\chi^{2}/d.o.f.$ \\

\startdata
non relat& 2.03 $\pm$ 0.05 & 101 $\pm$ 0.12 & - & - & 442.4/448 \\
non relat& 2.15 $\pm$ 0.07 & 93 $\pm$ 0.14 & 0 & $1.6^{+1.1}_{0.8}$ & 431.1/449 
\\
non relat& 2.11 $\pm$ 0.07 & 95 $\pm$ 0.14 & 500 &$ 0.45^{+0.6}_{0.3}$ & 436.4/4
49\\
relat    & 2.18 $\pm$ 0.08 & 91 $\pm$ 0.15 & $660^{+2400}_{-660}$ & 1.0 & 429.2/
449\\
relat    & 2.28 $\pm$ 0.07 & 85 $\pm$ 0.14 & 0 & $3.6^{+1.2}_{-1.7}$ &
425.1/449\\
\enddata

\tablenotetext{a}{Temperature in eV}
\tablecomments{The model consists of  black body, power law, Galactic
absorption (fixed value) and reflection component with an iron line.}
%
%\end{tabular}
%\end{table*}
\end{deluxetable}

\subsubsection{RXTE data - reflection component}

In order to put better constraints on the reflection parameters, 
we used the RXTE data, which nominally extend up to $\approx 20$ keV.  While
PG1211+143 in its flux state, as observed with the RXTE, is a relatively 
faint source to be studied in great detail by the RXTE's PCA, a
non-imaging instrument, the data still put some constraints 
on the crucial 8 - 20 keV band where the Compton reflection is
important.  We adopted the composite model consisting of a primary soft 
component (described as a Comptonized black body), power law continuum, 
and the Compton -- reflected spectrum with the iron $K\alpha$ line 
included (\. Zycki, Done \& Smith 1997;  1998).  The reprocessed spectrum 
is parameterized by $\xi$, $R=\Omega/2 \pi$, as above.  The outer disk 
radius, as before, was fixed at $10^4 R_g$ and the power law index 
describing the incident flux as a function of radius was fixed again at 
the value -3.0.  However, the inclination angle and the inner radius of the 
reflecting disk surface were now allowed to vary.   We only studied 
the relativistically smeared model since it is the most physical one.
The iron line energy, again, is not a free
parameter, but it depends on the ionization state of the reflecting 
medium.  Its profile is connected with the assumed geometrical 
parameters: inner radius of accretion disc and viewing angle.

\begin{figure}
\epsfxsize = 80 mm 
\epsfbox[10 100 600 700]{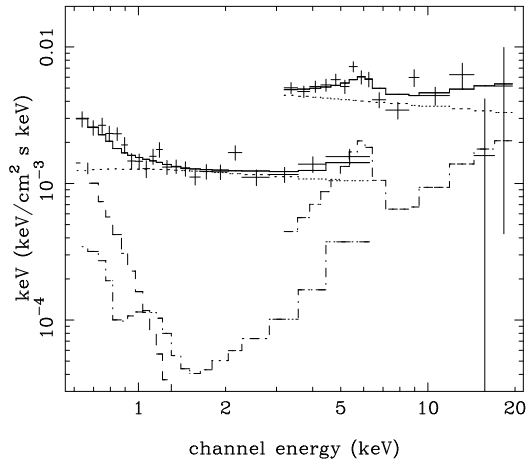}
\caption{The model fit (continuous histogram) to the combined non-simultaneous 
ASCA/XTE data (see Table ~\ref{fits_xte}). Dashed histogram shows the 
contribution of the soft {\sc comptt} component, dotted-dashed histogram - the
reflection component and dots mark the hard X-ray power law.  
\label{fig:ASCA/XTE}}
\end{figure}

The resulting best fit parameters are given in Table ~\ref{fits_xte} and
presented in Figure~\ref{fig:ASCA/XTE}.
New results did not differ significantly from the ROSAT/ASCA results 
regarding the soft part of the spectrum, although the optical depth of the 
Comptonizning medium dropped somewhat, and the hard X--ray power law 
became softer. However, they provide an interesting insight into 
the geometry. The amplitude of reflection is consistent with 1.0, 
and the reflecting area extends down to the marginally stable orbit 
which means that the disk is probably not disrupted. However, its surface
(at least in the inner parts dominating the reflection) is considerably
ionized ($\xi \sim 500$), and the neutral reflection is excluded at a 
$90 \%$ confidence level.  For such a case, the resulting iron line energy
corresponds to the domination of Fe XXII  ions and its rest frame
energy is $ E \sim 6.7$ keV, while the line equivalent width is $EW \sim 150$ 
eV. This supports the results of Yaqoob et al. (1994), who found a marginal
evidence for an ionized iron $K\alpha$ line fitting the ASCA data.  The 
inclination angle of the disk is low, i.e. we see the disk essentially 
face on. 

\begin{deluxetable}{lcccccccr}
\tablenum{7}
\small
\tablecaption{Results of simultaneous fitting to ASCA and RXTE data
\label{fits_xte}}
%\tablewidth{100 mm}
\tablehead{
\colhead{$\Gamma$} &
\colhead{ $kT_{soft}^a$} &
\colhead{ $kT^a$} &
\colhead{ $\tau$} &
\colhead{ $R$} &
\colhead{ $\xi$} &
\colhead{ $\cos i$} &
\colhead{ $R_{in}^b$} &
\colhead{ $\chi^{2}_{\nu}$} 
}

%\begin{table*}
%\caption{Results of simultaneous spectral fitting to ASCA and RXTE data. The model consists of the Comptonized black body, power law, ionized reflection with $K\alpha$ line and Galactic absorption (fixed value).
%\label{fits_xte}}
%\begin{tabular}{ccccc}
%$\Gamma$ & $kT_{soft}$  [keV]  & $kT$ [keV]  & $\tau$  & $\chi^{2}_{\nu}$ \\

\startdata
$2.18 \pm 0.03$  & $<0.03$ & $0.13 \pm 0.05$ & $12 \pm 
5$ &
$0.85 ^{+0.65}_{-0.55} $  & $500^{+600}_{-450}$ & $0.9 \pm 0.1 $ &
$6.0 2^{+4.0}_{-0.0}$ & 1.04 \\
\enddata

%\end{tabular}
%\end{table*}
\tablenotetext{a}{Temperature in keV}
\tablenotetext{b}{Inner radius in $R_{g}$}
\tablecomments{The model consists of  the Comptonized black body,
power law, Galactic
absorption (fixed value) and reflection component with an iron line.}

\end{deluxetable}

%{\bf The soft X--ray photon temperature is again not well constrained: firm upper limit is derived but the temperature can be arbitrarily low, apart from the limit
%provided by the model itself. This value has an important effect on the 
%spectral shape in EUV band, below 0.1 keV }.

\subsubsection{Optical/UV/X--ray data}

Finally, we include the optical/UV data into our modeling. 
For that purpose, we add a new model component, which essentially 
has a form of standard accretion disk of Shakura \& Sunyaev (1973) 
around a non-rotating black hole.  We assume that the disk, whether 
it is or it is not covered by the ionized skin, radiates locally as 
a black body. It is certainly a simplifying assumption (see e.g. 
Merloni, Fabian \& Ross 2000) but more accurate models critically 
depend on the assumed details of the disk vertical structure
(e.g. Nayakshin, Kazanas, \& Kallman 2000;  Madej \& R\'o\.za\'nska 
2000b).  Such assumptions are beyond the scope of the present paper. 

Fits of the reflected component suggest that the disk extends 
down to the marginally stable orbit.  However, the reflection 
component indicates a highly ionized hot disk surface, and thus 
we assume that the emission of the standard cold disk is only
detectable from radii larger than certain $R_{ion}$.  The
gravitational energy dissipated above $R_{ion}$ is directly 
re-emitted by the disk, while the energy dissipated below $R_{ion}$ 
has to provide the energy source for hard X--ray emission and 
Comptonizing medium.  

The disk parameters are therefore: the black hole mass $M$, the mass 
accretion rate $\dot M$, and $R_{ion}$. We assume a face-on view 
of the accretion disk, and the distance to the source calculated assuming 
$H_o = 75$ km s$^{-1}$Mpc$^{-1}$.

\begin{figure}
\epsfxsize = 80 mm 
\epsfbox{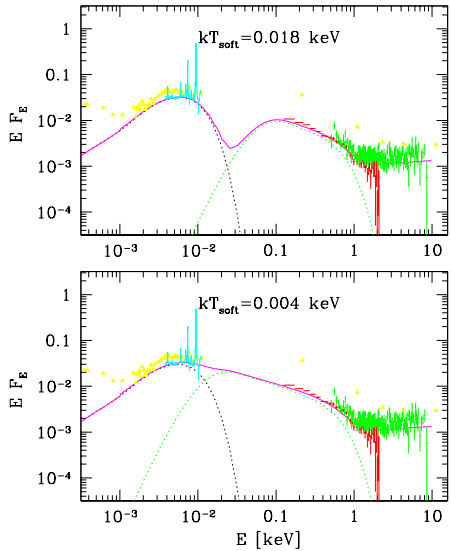}
\caption{The broad band spectrum of PG1211+143 in the IR - X range.  
In the lower panel we show the model  calculated 
assuming that the accretion disk provides seed photons for Comptonization by 
the hot plasma. In the upper panel we show the model which assumes that 
the soft photons are emitted in the hotter plasma.  
The solid triangles are data points from Elvis et al. (1994). 
The UV spectrum with emission lines  are the HST data (Bechtold 
et al. 2000) and the crosses are the X--ray data from ASCA and ROSAT.
The Comptonized disk  model was fitted to the X--ray data.  
In both cases, absorption in soft X--rays is  neglected in the plot, 
and for the modeling purposes, the X--ray data were corrected for absorption.
Optical/UV spectrum was modeled with a standard disk.  
Thick solid line is the sum of all model components.  
\label{spec2}}
\end{figure}

In the upper panel of the Figure ~\ref{spec2} we plot the broad band 
spectrum of PG1211+143, constructed for the case of the maximum value of 
temperature of the seed photons undergoing Comptonization. The solid 
line marks the sum of the cold accretion disc spectrum and the model 
fitted to the X--ray data that consists of the Comptonized black body 
(dashed line), power law function and Galactic absorption (which is 
neglected in the plot).

In this case there is a clear gap between the UV and soft X--ray band, 
which would be seen as a local minimum in the continuum around 
$2\times 10^{-2}$ keV.    
The bolometric luminosity measured between 10 000 \AA~ and 100 keV (assuming
an extrapolation of the measured hard X--ray power law) is 
$1.5 \times 10^{45}$ erg s$^{-1}$.
The mass of the black hole is $2.2 \times 10^8 M_{\odot}$, the 
accretion rate is 0.4$M_{\odot}$ yr$^{-1}$, and $R_{ion} = 28 R_g$. 
$L/L_{Edd}$ ratio in this solution is 0.05. About
50 \% of the energy is dissipated in the black body component of the
disk.  

In the lower panel of Figure ~\ref{spec2} we show the broad band 
spectrum determined under the assumption that cold disk itself 
provides the seed photons for Comptonization, i.e. the soft photon 
temperature is given by the effective temperature of the inner disk.
Such a spectrum forms a continuous Big Blue Bump extending from the 
optical/UV to soft X--ray band. The bolometric luminosity of the source is
in this case higher, $1.9 \times 10^{45}$ erg s$^{-1}$.
The mass of the black hole is $1.5 \times 10^8$ $M_{\odot}$, the 
accretion rate is 0.54 $M_{\odot}$ yr$^{-1}$, and $ R_{ion} = 42 R_g$. 
$L/L_{Edd}$ ratio in this solution is equal to 0.10.
About 30 \% of the energy is directly emitted by a 
black body outer disk, while 60 \% is in the Comptonized
component and $\sim 10$ \% in the hard X--ray power law. 

Both fits presented above reproduce the broad band spectral 
data well.  We note that all intermediate solutions are also 
possible and at a level of pure data analysis, we cannot reject 
any of the above possibilities.  
However, a composite spectrum of quasars given by Laor et al. (1997) and the
broad band spectrum of RE J1034+396 (e.g. Puchnarewicz et al. 2000) show 
clearly the existence of a single broad Big Blue Bump extending from 
optical/UV to soft X--ray band. Therefore using the analogy with these 
data we would favor our second solution.  

The mass of the central black hole determined from the spectral analysis 
is somewhat higher than the value determined from the
variability analysis (see Sect.~\ref{sect:masa}).  However, all 
quoted values are rather small for a quasar and comparable to the masses of 
the central black holes in Seyfert 1 galaxies such as NGC~5548, 
despite significantly higher bolometric luminosity. 

\section{Discussion}

\begin{figure}
\epsfxsize = 80 mm 
\epsfbox{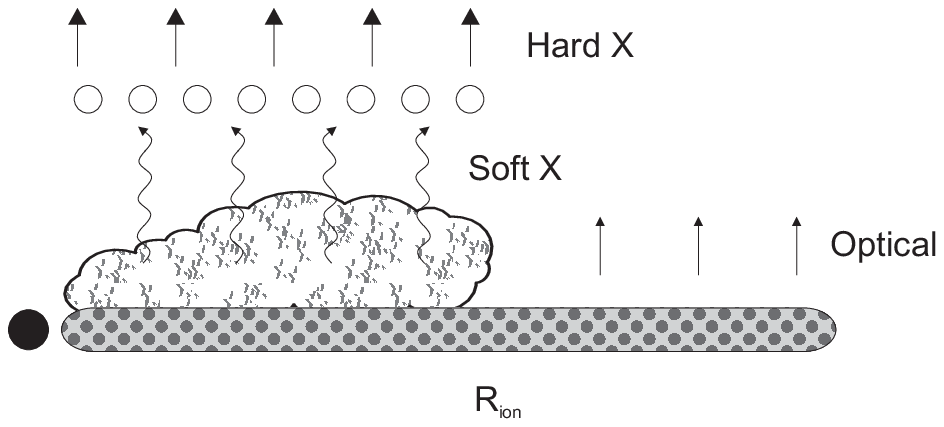}
\caption{The geometry of the accretion flow in PG1211+143 consistent with the
spectral model presented in Fig. \ref{spec2}. The optical flux is
emitted by the cold accretion disk ($T \sim 10^{4}$ K). The disk 
is the source of seed photons for the hot Comptonizing cloud 
($T \sim 10^{6}$ K, $\tau \sim 20$), which extends below the 
transition radius $R_{ion}$. The hard X--ray flux is emitted by the 
hot flare region ($T \sim 10^{9}$ K) and are partially reflected by the 
cloud ($\xi \sim 500$, $\Omega/2\pi \sim 1$).  
\label{fig:geom}}
\end{figure}

\subsection{Determination of the black hole mass}

The mass and luminosity estimates presented in this paper and those
available in the literature
render PG1211+143 as a relatively low-luminosity, low-mass object.
Quasars constituting
the sample of Laor et al. (1997) and Zhen et al. (1997) have black hole masses on average about
$1.4 \times 10^9 M_{\odot}$ but they have also higher luminosities, 
about $10^{46}$ erg s$^{-1}$. An 
accurate measurement, however, is quite difficult to obtain. 
Our method based on the X--ray
variability in the form of PDS measurement (see Sect. \ref{sect:masa}) 
gives $\log(M) = 7.0 \pm 0.7$, while the spectral fit indicates 
$\log(M) = 8.17$. We did not give the formal error of this 
second measurement since the systematic error certainly dominates 
any statistical error involved. Nonetheless, the error in the value 
of the mass determined from the spectral fitting can be roughly 
estimated under the assumption that the outer part of the flow is well
described by a standard Keplerian disk. The fit is based on two relationships.
The first one is the asymptotic shape of the disk emission in optical band
\begin{equation}
F_{\nu} \propto \nu^{1/3} (M \dot M)^{2/3} (\cos i) H_{o}^{2} f^{-4/3},
\end{equation}
where $\cos i$ is the inclination angle of the disk, $H_o$ is the
adopted value of the Hubble constant and $f$ is the spectral 
correction given by the color temperature to the effective temperature ratio.
The second constraint comes from the bolometric luminosity of the system
\begin{equation}
F_{bol} \propto \dot M \eta H_o^2,
\end{equation}
where $\eta$ is the efficiency of accretion. In our fits we adopted 
$H_o = 75$ km s$^{-1}$ Mpc$^{-1}$, $\cos i = 1.0$, $f=1$ and $\eta = 1/16$. However, if 
those values in reality are different from the assumed ones, 
the derived mass will scale as
\begin{equation}
M = 1.5 \times 10^8 f^2 \eta (\cos i)^{-1} (H_o/75)^{-1}.
\end{equation}

This means that the departure from a black body approximation and 
slight inclination of the system will tend to increase the mass, 
although not considerably.  The color to effective temperature 
ratio was determined to be $\sim 1.7$ (Shimura \&  Takahara 1995;  
Sobczak et al. 1999). Merloni, Fabian, \& Ross (2000) give even larger 
values for Galactic sources, but in the case of AGN, efficient bound-free 
transitions in heavy elements reduce $f$ significantly 
(R\'o\.za\'nska 2000, private communication).  Therefore 
the upper limit for $f$ seems to be about 1.7,  and the most 
probable inclination is $30^o$, giving together an increase in the
derived mass by a factor of 3. On the other hand, if there is a strong 
outflow close to the black hole, it will effectively decrease the standard 
efficiency of accretion, thus decreasing the mass. Strong 
outflow may give a factor of two, if it is comparable energetically
to the outflows in radio-loud objects. On the grounds of this analysis, 
we can estimate an error in our determination from the spectral analysis as 
$log(M) = 8.17 ^{+0.5}_{-0.3}$.  The lower limit is still not consistent 
within a formal error with the determination based on
variability, but in that case it is difficult to estimate the possible 
systematic error of the method. As for the determination of Kaspi 
et al. (2000), the formal errors are small: 
$4.05^{+0.96}_{-1.21} \times 10^7 M_{\odot}$,  but 
recent analysis of the systematic errors  of methods based 
on BLR properties indicates a possible factor of three error in 
both directions, due to technical problems with sampling as well
as a possible (and quite probable) flattening of the BLR (Krolik 2000). 

All this implies that taking the systematic errors into account, all 
measurements are roughly consistent with each other, even though they are 
highly inaccurate.  We notice, however, that if we allowed for a 
maximally rotating black hole as a central object, our mass based
on the spectral fits would be higher by a factor of 7. This result,
however, would be rather difficult to reconcile with other measurements.

\subsection{Luminosity state}
PG1211+143 is both a quasar, and a Narrow Line Seyfert 1 galaxy. Temporal and
spectral analysis presented in Section~\ref{sect:results} show
that the mass of the central black hole is rather low ($\sim 5 \times 10^7 - 
2 \times 10^8 M_{\odot}$) for a source of such optical/X--ray bolometric 
luminosity ($\sim 1.5 - 1.9 \times 10^{45}$ erg s$^{-1}$). In a sample of 
Kaspi et al. (2000) it is on the border between normal Seyfert galaxies,
NLS1 and quasars, having mass typical for a Seyfert1 but narrow $H_{\beta}$
line and larger luminosity than normal Seyferts. An interesting comparison
can be made with Fairall 9 galaxy having broad $H_{\beta}$: when PG1211+143
is faint and Fairall 9 is exceptionally bright (Clavel, 
Wamsteker \& Glass 1989) their luminosities are nearly equal.
For those cases, the luminosity in
Eddington units $L/L_{Edd}$ of PG 1211+143 remains only marginally larger than 
for Fairall 9 (we adopted the black hole mass of Firall 9 to be  twice that of PG1211+143, after Kaspi et al. 2000).

The exact value of $L/L_{Edd}$
is not determined accurately since it depends on the black hole mass and 
bolometric luminosity determination. 
Our
$L/L_{Edd}$ ratio determined from spectral fits ($\sim 0.05 - 0.1$) 
is somewhat lower than the value of 0.29 given by Kaspi et al. (2000) due to 
larger black hole mass and lower bolometric luminosity than resulted from
BLR study and the use of 
monochromatic luminosity at 5100 \AA.
The source is strongly variable and our fits (as well as results of Kaspi et al. 2000) are
mostly representative for a relatively faint state.

This result supports the view
expressed by many authors (e.g. Pounds et al. 1995;  Pounds \& Vaughan 2000) 
that Narrow Line Seyfert 1 galaxies are objects accreting at a relatively high 
$L/L_{Edd}$ ratio.  The analysis of the Compton reflection implies
that the inclination to the accretion disk surrounding the 
nucleus is low, which is contrary to the scenario where NLS1 
galaxies are viewed at a high inclination and their variability is caused by
obscuration events (Brandt \& Gallagher 2000;  see also Boller 2000).

Weak hard X--ray producing corona (as compared to the large luminosity 
of the Blue Bump) and large $L/L_{Edd}$ ratio are consistent 
with the prediction of the two-temperature corona model (Janiuk \& Czerny 
2000), and the relatively low photon index even in this source 
($\Gamma = 2.18$) suggests that models with strong advection cannot
reproduce the observed spectrum well  (Janiuk, \. Zycki, \& Czerny 2000).
However, a non-accreting magnetic corona model, perhaps including 
some clumpiness, may well be viable, as argued by Poutanen \& Fabian (1999). 

\subsection{Reflecting and Comptonizing medium}

Our spectral analysis requires the existence of a number of various 
plasma zones: (1) very hot plasma, responsible
for the hard X--ray emission, (2) moderately hot, moderately dense 
plasma responsible for Comptonization, (3) moderately hot 
plasma responsible for reflection and (4) a cold, dense 
disk. It is natural to pose the question whether the components (2) and
(3) are actually the same medium but parameterized differently in the
spectral fits.

Medium (2) has the temperature $\sim 0.15$ keV, or $10^6 $K, and a 
large optical depth, $\tau \sim 10$. Medium (3) has the ionization 
parameter $\xi \sim 500$.  Temperatures intermediate between the 
black body temperature and the Inverse Compton temperature are 
difficult to achieve by radiatively heated plasma, since there is 
a rapid transition between the two solutions (Krolik, McKee \& Tarter 1981).   
This transition happens at $\Xi \approx 10$ (see, e.g., \. Zycki \& 
Czerny 1994), and the relationship between the two ionization parameters is
$\xi = 5\times 10^{-5} T \Xi$.  Plasma in zones (2) and (3) should be 
in such a transition state.  Assuming $\Xi=10$ and $\xi=500$ from
plasma zone (3) we obtain the temperature consistent with the
temperature of plasma zone (2). This would imply that the same medium 
which Comptonizes the disk photons coming from below also reflects 
the hard X--ray photons illuminating it from above.  

Such interpretation clearly simplifies the geometry of the flow. The
accreting stream consists simply of the outer black body disk and an 
inner disk which develops very optically thick ($\tau \sim 10$), warm skin 
($T \sim 10^6 $K). A small fraction of energy is released in the form 
of hard X--ray emission above this skin.  A possible schematic view of 
the flow is shown in Figure~\ref{fig:geom}.

Similar values of the optical depth and the temperature of the
``warm'' plasma were derived for NGC~5548 by Magdziarz et al. (1998). 
However, in this source the magnitude of reflection was lower and it 
was not possible to locate geometrically the 
Comptonizing medium. Magdziarz et al. (1998) actually suggested that the radial
transition zone from a classical disk to an inner ADAF-like flow is the most
probable location. Another example of the presence of the warm skin
was inferred from the XMM observations of PKS 0558-504 (O'Brien et al. 2000).
Interestingly, numerous examples of a Comptonizing medium that 
modifies the shape of the spectrum in soft X--ray band were inferred 
from data for Galactic sources:  for those, even before the effects of
Comptonization, this soft component extends 
to higher energies as compared with AGN due to the higher disk 
temperature, and therefore can be more easily observed. Examples 
include Nova Muscae 1991 (\. Zycki et al. 1998), Cyg X-1 
(Gierli\' nski et al. 1999), GRS 1915+105 (Vilhu \& Nevalainen 1998), 
GS 2000+25, XTE J1550-564, GRO J1655-40 (\. Zycki et al., in preparation), 
and the problem has been discussed in a review of \. Zycki (2000).

Our results seem to indicate that, at least in sources like PG1211+143, this
warm medium forms a horizontal layer. Its optical depth is only 
$\tau \sim 10$, which is rather low in comparison to the total 
optical depth of the classical disk.  Still, about 40 - 60\% of the total 
gravitational energy dissipated by the flow is released in this 
zone, while the remaining $\sim 50 - 30 \%$ is released in the 
body of the disk.  Roughly 10\% of energy is emitted in hard X--rays
and thus presumably released in the hot corona.  However, this
quantity is difficult to estimate due to the unobserved high energy cutoff. 

\subsection{Formation of the warm skin}
\label{warm_form}

Application of the thermal instability of the irradiated hot plasma by Krolik
et al. (1981) to the vertical structure of the disk clearly shows that a warm
skin, or a corona at the Inverse Compton temperature, forms above the cold 
disk.   However, the details of the sharp transition between the disk and the
Inverse Compton -- heated corona are not clear (e.g. 
Begelman, McKee, \& Shields 1983; R\' o\. za\' nska \& Czerny 1996;
2000; Nayakshin, Kazanas, \& Kallman 2000;  Nayakshin 2000).  

Current models imply that the optical depth of this transition zone 
of intermediate temperature is generally very low and the total optical 
depth of the heated skin is never large.  Results of Nayakshin et al. 
(2000) and \. Zycki \& R\' o\. za\' nska (2000) allow at most for 
$\tau \sim 1.5 $ even if the X--ray radiation is released in compact 
regions leading to an enhancement of the irradiation under a hot spot 
by as much as two orders of magnitude.  This is simply related to the fact thatthe X--ray heating does not penetrate deeply enough to support the high 
temperature of the transition zone.

Since the Comptonized component contains a significant fraction of the 
bolometric luminosity ($\sim 40 - 60$ \%), much larger than the fraction 
contained in the hard X--ray part (up to $\sim 10$ \%) the Comptonizing medium
cannot be heated by hard X--rays.  For this reason, 
the discrepancy between the results quoted
above and the large optical depth determined from our fits is not surprising.
The dissipation must occur within this zone and we can attempt to compare
the required amount of heating with most popular parametric prescription 
for the viscous dissipation of gravitational energy.

The properties of the reflection and Comptonization by the warm skin allow 
us to estimate the number density and the temperature of this medium at the 
optical depth of $\tau \sim  1.0$.  The temperature $T_{warm} \sim 10^6 $ K 
results directly from the {\sc comptt} fit to the data.  The density 
can be determined from the ionization parameter of reflection $\xi =500$, 
assuming that the typical radius under discussion is $20 R_g$ and the 
bolometric luminosity in hard X--ray band is $10^{44}$ erg s$^{-1}$:  
\begin{equation}
n= {L_X \over r^2 \xi} \sim 10^{12}~ {\rm cm}^{-3},
\end{equation}
or equivalently, $\rho \sim 10^{-12}$ g cm$^{-3}$.
This medium is clearly Compton-cooled, since the efficiency of bremsstrahlung 
for such parameters, namely 
\begin{equation}
\Lambda_B = 6.6 \times 10^{20} \rho^2 T^{1/2} \sim 0.006~ {\rm erg~ s}^{-1} {\rm
 cm}^{-3}
\end{equation}
is two orders of magnitude lower than efficiency of Comptonization, which is  
\begin{equation}
\Lambda_C = F_{tot} \kappa_{es} \rho {4 k T \over m_e c^2} \sim 0.2~ {\rm erg~ s
}^{-1} {\rm cm}^{-3},
\end{equation}
where the value of 
$F_{tot} \sim 10^{15} {\rm erg s}^{-1} {\rm cm}^{-2}$ 
was estimated from the total bolometric luminosity $10^{45}$ erg
s$^{-1}$ and the distance of $20 R_g$.

This large amount of cooling has to be compensated by the heating.
The most popular parameterization of the heating in accretion flow is
the $\alpha$ viscosity prescription introduced by Shakura \& Sunyaev (1973).  
Two variants are most popular: $\alpha P_{tot}$ and $\alpha P_{gas}$.  
In the first case the amount of heating (in the context of a Keplerian
flow) is given by
\begin{equation}
Q^+_{Ptot} = 1.5 \alpha P_{tot} \Omega_K \sim 0.3 \alpha~~ {\rm erg~ s}^{-1} {\rm cm}^{-3},
\end{equation}
where we approximated $P_{tot}$ by $P_{rad} \sim F_{tot}/c$;  the
Keplerian angular velocity $\Omega_K$ was estimated at $20 R_g$
for a $10^8 M_{\odot}$ black hole mass.
In the second case
\begin{equation}
Q^+_{Pgas} = 1.5 \alpha P_{gas} \Omega_K \sim  3 \alpha \times 10^{-3}~~ {\rm er
g~ s}^{-1} {\rm cm}^{-3},
\end{equation}
where the density and the temperature were taken from the previous estimates.

We see that the $\alpha P_{gas}$ variant of heating is by two orders 
of magnitude too small to compensate for the cooling.  However, 
the $\alpha P_{tot}$ variant is of the right order of magnitude, 
provided the viscosity coefficient $\alpha$
is  large. It does not explain the physical mechanism behind
the heating but clearly indicates that $\alpha P_{tot}$ scaling is
a good phenomenological description of the process.

\subsection{Consequences of the $\alpha P_{tot}$ heating}

There are two major consequences of the scaling favored by our analysis 
for the dissipation:  disk instability and surface temperature gradient. 

The disk models are known to be unstable under such scaling if the radiation 
pressure dominates, as it is the case in disks in active galaxies 
(Pringle, Rees, \& Pacholczyk 1973; Ligthman \& Eardley 1974). The disk is
supposed to display strong outbursts in timescales of years, with about two 
orders of magnitude difference between the low and high luminosity
states, as it is suggested by direct scaling of computations of 
the time evolution of Galactic sources (e.g. Honma et al. 1991;  
Szuszkiewicz \& Miller 1998;  Janiuk et al. 2000). 

The variations of the Big Blue Bump on these timescales are observed
in active galaxies in general 
(see Sect.~\ref{sect:long}), and the maximum variability (peak on PDS
diagram) has not been determined yet for the far UV and soft X--ray
band.  This may be because such a peak is possibly at even 
longer timescales than can be inferred from observations covering a
span of only 20 years.  
Therefore, we cannot yet determine the viscosity parameter $\alpha$ governing
the timescales of outbursts.  One aspect of the current 
data which may be in conflict with the above scenario is that the observed
variations of the source luminosity do not appear as a fraction of
a global outburst with the predicted large amplitude. There are, however,
various effects which may lead to the difference between the modeled outbursts
and the observed time behavior. The models studied so far did not include 
a warm skin as that 
discussed in Sect.~\ref{warm_form}. Moreover, any strong outflow 
is known to suppress the outburst amplitude, if the outflow is coupled to 
the temporary local accretion rate (Janiuk et al. 2000), and any 
irregularities in this outflow lead immediately to much more complex 
lightcurve than simple predictions would indicate 
(Nayakshin, Rappaport, \& Melia 2000).
In addition, if the viscosity parameter depends on the amount of the radiation 
pressure, the amplitude of the outburst can be reduced.  Testing such scenarios
is more promising in case of Galactic sources because of the shorter timescalesrequired. 

Another aspect of the scenario that may present a conflict with
observations is connected with the broadly discussed issue of the
lack of Lyman edge in the observed spectra of AGN while the presence of such an
edge is predicted by most models of accretion disk atmospheres (for a review, see Koratkar
\& Blaes 1999). Those studies, 
however, never included such a strong heating close to the surface as well
as the accompanying Comptonization. Strong heating leads to a more isothermal
atmosphere, as already argued by Laor \& Netzer (1989), and 
Comptonization is also quite efficient in removing the edge, as stressed
by Czerny \& Zbyszewska (1991). Some of the recent papers already made a step
towards reducing the predicted edge (R\' o\. za\' nska et al. 1999;  
Madej \& R\' o\. za\' nska 2000a;  Hubeny \& Hubeny 1998) 
but most likely the issue
cannot be fully resolved until heating is properly included in computations 
of the disk atmosphere.  

\subsection{Future observational tests}      %(jako 4.5)

High resolution spectroscopy in soft X--ray band  available now through 
{\sl Chandra} and XMM data may provide an independent test of the physical
conditions in the reflecting/Comptonizing warm skin. This region
is not completely ionized so it is a source of intense emission lines, mostly
of the hydrogen-like Ly$\alpha$ lines from CNO as well as Si, Mg, S and Ne.
The presence of such lines has been possibly detected in the 
MCG--6-30-15 spectrum taken with the XMM 
(Branduardi-Raymont et al. 2000).

Computations of the irradiated slab predict the existence of such lines
(Ross, Fabian, \& Young 1999). Dumont et al. (2000) performed the 
radiative transfer in an irradiated slab under constant pressure, 
with the spectral slope of the incident radiation $\Gamma = 2.0$ 
and the high energy cut-off of 100 keV.  In this case
the equivalent width of the Ly$\alpha$ CVI, NVII and OVIII lines were
equal to 1.6 eV, 2.3 eV and 7.4 eV for the ionization parameter $\xi$ (at the
surface) of 300, and 1.3 eV, 2.8 eV and 11.5 eV for $\xi = 1000$.
Those values are quite similar to the EW measured 
in the X--ray spectrum of MCG--6-30-15 (Branduardi-Raymont et al. 2000).
However, since in those considerations the slab was only radiatively heated
from the outside, the high temperature zone was relatively thin (Thompson
depth of 0.1 and 0.3, correspondingly). The optical depth of the Comptonizing
medium determined by the data was on the order of 10.  With this, it
is difficult to predict if the additional heating in the interior of the 
warm skin interior may change these results.

On the other hand, the theoretical computations performed for a constant 
density irradiated medium, with harder incident flux ($\Gamma = 1.9$, 
extending up to 280 keV) indicated much larger EW, of order of 50 eV for
$\xi = 1000$ (Czerny \& Dumont 1998) which suggests that EW of the soft 
X--ray lines is sensitive to the structure of the 
warm skin. Clearly, more detailed
theoretical studies together with observational constraints on the line
intensity and its profiles may help to resolve the problem of the
formation of such warm skin. The models should include both the effect of 
Comptonization and kinematical/gravitational broadening of the line
profiles, as both processes lead to development of red wings in
emission lines, and may be difficult to differentiate. 

\section{Conclusions}

The Narrow Line quasar PG1211+143 is considered to be a very good 
candidate to study the broad band emission and variability properties
due to a rich sample of observational data available for this object.
In this work, we considered the physical mechanisms which may be
responsible for the optical, UV and X--ray emission and variability
of this source, and we found that the following model can give a 
satisfactory explanation of the observations:
\begin{itemize}
\item The central object powering the source is a supermassive black
hole with a mass 
$M = 10^{7} - 2\times 10^{8} M_{\odot}$ and the luminosity to
Eddington luminosity ratio is in the range of 5 - 40 \%. Despite
large errors both quantities are consistent with the explanation of
the phenomenon of Narrow Line Seyfert galaxies based on a relatively low mass
of the black hole and a high accretion rate.
\item The emitted radiation originates in the cold accretion disk
responsible for the optical/UV spectrum. Inner disk region develops a
warm, optically thick skin, which produces the profound ``soft
excess'' observed in the X--ray spectrum and is well modeled with a
Comptonized black body emission. However, from the spectral
analysis alone, it cannot be determined whether the source of seed photons
for Comptonization is the cold accretion disk or a somewhat hotter
plasma, and both explanations are possible.
\item The warm disk skin is also responsible for the reflection of
hard X--rays, which are emitted in the accretion disk corona. The latter has
 the form of active flares rather than forming a continuous medium.
\item Both accretion disk and its warm skin are heated by the viscous 
dissipation process, which can be parameterized by $\alpha P_{tot}$
prescription. This may lead to the source luminosity variations
resulting from 
the presence of regions where the radiation pressure is dominant.
\item Measurements of the soft X--ray emission lines possible in 
the {\sl Chandra} and 
XMM data and their comparison with theoretical models may help to 
test the warm skin structure.

\end{itemize}

%%\section{Acknowledgments}

\medskip

{\sl Acknowledgments.}  We thank Piotr \. Zycki and Asia Kuraszkiewicz for 
helpful discussions, and to Jill Bechtold, Adam Dobrzycki and their coauthors
for supplying us with their HST spectrum of PG121+143 prior to publication.  
In the course of our research, we used data obtained 
through the HEASARC on-line service, provided by NASA/GSFC. This work
was supported in part by grant 2P03D01519 of the Polish State
Committee for Scientific Research. GM acknowledges support 
by NASA RXTE Observing Grant via the ADP program.

\clearpage
\end{document}